# Sunspot Drawings by Japanese Official Astronomers in 1749-1750

Hisashi Hayakawa (1, 2)*, Kiyomi Iwahashi (3), Masashi Fujiyama (4), Toshiki Kawai (4), Shin Toriumi (5), Hideyuki Hotta (6), Haruhisa Iijima (4), Shinsuke Imada (4), Harufumi Tamazawa (7, 8), Kazunari Shibata (7)

(1) Graduate School of Letters, Osaka University, 5600043, Toyonaka, Japan (JSPS Research Fellow).
(2) Rutherford Appleton Laboratory, Chilton, Didcot, Oxon OX11 0QX, UK
(3) National Institute of Japanese Literature, 1900014, Tachikawa, Japan
(4) Institute for Space-Earth Environmental Research, Nagoya University, 4640814, Nagoya, Japan
(5) National Astronomical Observatory of Japan, 1818588, Mitaka, Japan (NAOJ Fellow).
(6) Graduate School of Science, Chiba University, 2638522, Chiba, Japan
(7) Kwasan Observatory, Kyoto University, 6078471, Kyoto, Japan
(8) Disaster Prevention Research Institute, Kyoto University, 6110011, Uji, Japan.

*email: hayakawa@kwasan.kyoto-u.ac.jp

**Abstract**

Sunspot observations with telescopes in 18[th] century were carried out in Japan as well. One of these sunspot observations is recorded in an account called *Sansaizusetsu narabini Kansei irai Jissoku Zusetsu* (*Charts of Three Worlds and Diagrams of Actual Observations since Kansei Era*). We analyze manuscripts of this account to show in total 15 sunspot drawings in 1749-1750. These observations were carried out by contemporary official astronomers in Japan, with telescopes covered by *zongurasu*s (< *zonglas* in Dutch, corresponding to "sunglass" in English). We count their group number of sunspots to locate them in long-term solar activity and show that their observations were situated around the solar maximum in 1749 or 1750. We also computed their locations and areas, while we have to admit the difference of variant manuscripts with one another. These observational records show the spread of sunspot observations not only in Europe but also in Japan and hence may contribute to crosscheck or possibly improve the known sunspot indices.

## 1. Introduction

Solar activity is reviewed and measured by appearance of sunspots on the solar disc (Vaquero & Vázquez, 2009). It is in the early 17$^{th}$ century that telescopic observations for sunspots started to offer scientific datasets from one of the longest-running experiments in human history (Owens, 2013). These sunspot observations were used to reconstruct past solar activity. R. Wolf and his successors in Zürich construct the Zürich number (Wolf number) to cover solar activity since 1700 (Waldmeier, 1961). Hoyt and Schatten (1998) include more observations to construct group sunspot number since 1610s.

Recent studies revisit the sunspot number (Clette *et al.*, 2014; Svalgaard and Schatten, 2016; Vaquero *et al.*, 2016; Willamo *et al.*, 2017), based on recent discussions on original sunspot drawings and sunspot counting within scientific documents after telescopic observations since 1610s (*e.g.*, Vaquero, 2007; Vaquero and Vázquez, 2009; Arlt, 2008, 2009, 2011; Arlt, 2009; Arlt and Fröhlich, 2012; Pavai et al., 2016; Cliver and Keer, 2012; Diercke *et al.*, 2014; Usoskin *et al.*, 2015; Willis *et al.*, 2013, 2016a, 2016b; Arlt *et al.*, 2016; Carrasco *et al.*, 2016, 2018; Hayakawa *et al.*, 2018a, 2018b; Svalgaard, 2017). Within these datasets, sunspot drawings are of greater value as they have information not only on sunspot number but also on their area, distribution, locations, configuration, and so forth (Vaquero, 2007; Vaquero and Vázquez, 2009).

While most of the early sunspot drawings down to the end of 19$^{th}$ century were from Europe (Vaquero and Vázquez, 2009; Vaquero *et al.*, 2016), recent findings of early sunspot drawings from non-European countries contribute to improve the reconstruction of past sunspot index (*e.g.*, Domínguez-Castro *et al.*, 2017; Denig and McVaugh, 2017). The Japanese archives may also contribute to this reconstruction by their contemporary sunspot drawings. Hoyt & Schatten (1998) seem partially aware of sunspot drawings by Kunitomo Ikkansai (國友一貫齋)[1] during 1835-36 (Yamamoto, 1937; Kubota & Suzuki, 2003). However, even before his sunspot observations, it is partially mentioned that we had some sunspot observations in Japan in 18$^{th}$ century (Kanda, 1960; Watanabe, 1987). One of them is sunspot drawings in manuscripts of *Sansaizusetsu narabini Kansei irai Jissoku Zusetsu* which can be translated as *Charts of Three Worlds and Diagrams of Actual Observations since Kansei Era* (三際圖説 並寛政以來實測圖説, hereafter, SKJZ). In this article, we examine SKJZ to show their sunspot drawings and relevant records with their digitalization. We also count their group number of sunspots and compare them with contemporary solar activity.

---

[1] In this article, we show Japanese personal names in orders of family name and first name as seen in the contemporary historical documents.

## 2. Method

SKJZ has two manuscripts with sunspot drawings (KS[2]: v.3, p.782). Both of them are preserved in Tohoku University at Sendai, Japan as shown below. We show their references with abbreviation, reference number, and hosting library.

MS/K: MS 8-21318-1 in Kano Library of Tohoku University Library

MS/O: MS 911-17799 in Okamoto Library of Tohoku University Library

We first introduce the characteristics of these manuscripts and show their sunspot drawings. We then analyze their text to estimate the observer and observational method. Then we count their group number of sunspots to locate them in the records to long-term solar activity. We also scaled the area of sunspots for their projected area and corrected area, before and after removing foreshortening effects. In this counting, we grouped sunspots according to the Zürich classification (e.g. Waldmeier, 1947; Kiepenheuer, 1953).

## 3. Result and Discussion

### 3.1. Manuscripts

As described above, we have two manuscripts of SKJZ with sunspot drawings. Colophons of both manuscripts relate their author or compiler with Watanabe Masanami (渡部将南), although his name is not found in other contemporary documents. MS/K consists of 31 folios while MS/O consists of 27 folios. Their difference is found in additional folios of MS/K (ff.27b-31a)[3] for drawings of comet observations on 1811 September 15, 1819 July 17, 1824 January 02, and 1825 October 03. On contrary, we can find two graffiti in red letters in MS/O (ff.2a-2b).

These manuscripts start with *Sansaizusetsu* to explain three phases from the ground to the upper sky generating thunder in Japanese traditional understanding (ff.1a-4b) and cover astronomical and meteological observations such as comets, sunspots, and solar halos from 1758 to 1825. We estimate MS/K and MS/O compiled at least after 1825 and 1803 according to their last date of observations: comet drawing dated 1825 for MS/K and drawing of halo dated 1803 for MS/O. These manuscripts include considerable amount of earlier observations copied from preceding sources including Watanabe Masanami's works.

---

[2] *Kokusho Soumokuroku*: a union catalogue for early Japanese books. Note that we have two other variants entitled as *Sansaizusetsu* (*Charts of Three Worlds*) without *Kansei irai Jissoku Zusetsu* (*Diagrams of Actual Observations since Kansei Era*). As sunspot drawings are included in the section of *Kansei irai Jissoku Zusetsu*, these variants do not involve sunspot records.

[3] We show the references with their folio number that is shown as f. for a single folio or ff. for plural folios.

We estimate the editor of this work as someone close to Shogunate Observatory as they involve several Shogunate observations in 1749-1750 or 1769-1770 (ff.8b-10a). One of the possible contributors is Toita Zentaro (戸板善太郎, 1708-1784) found in a correspondence to Yamaji Family in Shogunate Observatory (ff. 12a-12b). Hence, we estimate their editor as some of successors of Toita Zentaro in Observatory of the Sendai Fief (仙台藩). The fact that these manuscripts are preserved in the Tohoku University after the Sendai Fief also supports our estimation.

### 3.2. Sunspot Drawings

The sunspot drawings are found in ff. 8b-10a in MS/K and in MS/O as shown in Figures 1 and 2. Each manuscript involves 1 sunspot drawing (S1) with explanation of sunspot observation (f. 9a) and 14 sunspot drawings (S2-S15) with observational dates in traditional luni-solar calendar. We count their group number and summarize them with their date and local time (LT) converted to that in Gregorian calendar based on the conversion table (Uchida, 1992) and reference in Table 1. The explanation for S1 is transcribed and translated as follows.

> Original Text: 一 同年六月廿五日申半刻、又同廿六日辰刻前并午刻過ト申刻過見之、日体ノ内ニ黒点有、西洋暦経ニ所記無疑也、又日光熾也ト雖トモ輪ノフチキツハリト見ヘ月輪ノキワヲ見ルト異也
>
> English Translation: In the same year (1749), we observed the black spots within the solar disc on August 07 around 16h LT and on August 08 around 7h LT, 12h LT, and 16h LT[4]. This is without doubt as found in *Seiyo Rekikyo* as well. Although the sunlight is intense, the edge of the solar disc is clearly seen and different from that of the lunar disc.

Here, we can find the sunspot observational time and their knowledge on the solar disc. S1 (f.9a) involves the sunspot drawings on 1749 August 07-08 in one figure. The other sunspot drawings found in ff.9b-10a are drawn independently with one another as summarized in Table 1. We also identify the sunspot groups except for that of S1 in these manuscripts as shown in Figure 3.

Their observational time is described as 12h LT (午中) for each. These drawings reflect not only location and number of sunspots but their area and shape. For example, we find relatively large sunspots in S6 (1749 October 21) and S7 (1749 October 23). From the series of sunspot drawings, it is possible to roughly align the solar coordinates as well. We interpret the comment "upper side (上)" and "downer side (下)" on S2 defines the direction of panels and indicates that the drawings are placed such that north at local noon is always up. Note that nevertheless further finding of relevant sunspot drawing may change this interpretation, as this document does not provide further detail. Also, the evolution of the spot distribution in the three consecutive drawings from S7 to S9 hints that the Sun's rotation is from the lower left to the upper right (i.e., the rotational axis is oriented from the lower right to the upper left) and thus the drawings reflect what the Sun looks like and are not mirror symmetric. Note we assume that the large spot in the three drawings is identical and this spot rotates over the disk since it is unlikely that a spot of this size decays and emerges again in such a short time (4 days). If fact, the P angle in this period (October 1749) is +25 to +27 degrees, which means that the rotation axis for these panels is tilted counter-clockwise from the vertical by 25 to 27 degrees. Nevertheless, we have to admit that carefully comparing manuscripts MS/K and MS/O tells us that these drawings seem copied from some original drawings and the shape and size of every sunspot seems only slightly different. Therefore, we compute the spot locations for both manuscripts, which are summarized in Tables 2a and 2b.

---

[4] These observational clock hours are reproduced as the caption of S1 with its "upper side (上)" shown. Nevertheless, it is not described how these "clock hour" are computed here in this document.

### 3.3. Observational Instruments and Background Knowledge in 1749-1750

The observations in 1749-1750 including the sunspot observations in question seem carried out at the Shogunate observatory. These manuscripts (f.8b) explicitly state that the observations were carried out at *Bukou no Sokuryou Goyoujo* (武江之測量御用場), *i.e.* the Shogunate observatory located at current Sakuma-cho, Kanda in Tokyo (N35°42′, E139°47′) (Watanabe, 1987).

As for telescopes, we find descriptions of the great glasses (大御眼鏡) or the great telescopes (大遠鏡) without further details (ff.8b-9a). It is well known that contemporary high lord (将軍) Tokugawa Yoshimune (徳川吉宗, r.1716-1745, d.1751) was interested in astronomy and scientific technology in the contemporary Europe. He seems not content with astronomical instruments imported from Europe and even ordered to make some new instruments such as astrolabes for his own observation (*Tokugawa Jikki*, v.9, pp.292-293). It is also known that a special instrument called *zongurasu* (ぞんぐらす < *zonglas* in Dutch) was used as a filtering glass to enable observers to see the sun, got imported from the Netherlands at least until his death in 1751 (*Tokugawa Jikki*, v.9, p.294). The term *zongurasu* is a loan-word of *zonglas* for telescopes in Dutch (Vos, 2014). The modern Dutch also have a word "zonneglas" which means "a dark or soot-covered glass used for observation of the Sun" according to the comprehensive dictionary by van Dale (p.3535). The term *zonglas* is also found for a contemporary eye-tube number 35 made in 1723 within the catalogue for telescopes in the Leiden Observatory during 1656-1859 (Zuidelvaart, 2007, p.166). Therefore, we can fairly estimate the contemporary observers in the Shogunate Observatory covered telescopes with *zongurasu*s (< *zonglas*) probably immediately after getting them from the Netherlands.

Their knowledge on sunspot is also influenced by contemporary European astronomy. These manuscripts explicitly mention that they have studied about sunspots from *Seiyo Rekikyo* (西洋暦経). This book is considered an original book of *Rekisan Zensho* (暦算全書), a Chinese monograph introducing European astronomy and almanac by Méi Wéndǐng (梅文鼎) as described in *Tokugawa Jikki* (v.9, p.292).

This fact tells us that the Shogunate astronomers got considerable influence from European Astronomy via China. On contrary, the Shogunate astronomers seem to learn from ancient Japanese historical documents as well to adapt terminology of "black spots (黒点/黑點)" found in Japanese Official History in 851 (Hayakawa *et al.*, 2017b), while sunspots in Chinese historical documents are described in different character (黒子) (*e.g.* Keimatsu, 1970; Hayakawa *et al.*, 2015, 2017a, 2017b, 2017c; Tamazawa *et al.*, 2017).

### 3.4. Location in Long-Term Solar Activity

We count the group number of sunspots of the sunspot drawings in SKJZ during 1749-50, as shown in Table 1. While we have only 15 sunspot drawings here, we can locate these sunspot drawings in longer-term sunspot activity. Clette *et al.* (2014) offer us yearly sunspot number since 1611 and monthly sunspot number since 1749. These sunspot drawings are contemporary with the earliest stage of the monthly sunspot number compiled by Clette *et al.* (2014). Figures 4 and 5 show where these sunspot drawings are located in comparison with yearly sunspot number and monthly sunspot number by Clette *et al.* (2014). They show that these sunspot drawings during 1749-50 are near the solar maximum. This fact explains why there are relatively large number sunspots captured in SKJZ. We also find sunspots with large area in S6 and S7 on 1749 Oct 21 and Oct 23 as well. They are of great value as they can fill the unrecorded sunspot observations as shown in Table 1 in comparison with Vaquero *et al.* (2016).

We also computed area of each sunspots in these manuscripts identified in Figure 3, as shown in Tables 2a and 2b and summarized in Figure 7. We calculate projected area and corrected area, before and after removing foreshortening effect, for each manuscript and show their variation in Figure 7. We also show the histogram of sunspot area in projected area and corrected area for both manuscript. Figures 7 and 8 explicitly show that the sunspot areas in MS/O are relatively smaller than those in MS/K, and hence we need to be careful to discuss distribution of sunspot areas. These figures show that sunspot areas in these manuscripts are mostly up to 2000 msh except for S07_E01 in projected area and corrected area, and hence agrees with that of the modern spots (e.g. Hathaway *et al.*, 2002). S07_E01 is in somewhat extraordinary value of 3500 msh in MS/O and 4700 msh in MS/K in corrected area, while they are 2462 msh in MS/O and 3705 msh in MS/K in projected area (see, Aulanier *et al.*, 2013; Toriumi *et al.*, 2017). This somewhat extraordinary value may have been caused by its location near the solar limb. It is possible to consider the observer or copyist(s) mistakenly draw sunspot in the relatively similar shape with that of S06_E01, although it got closer to the solar limb and should have gotten foreshortening effect in its appearance. While we are not sure about the accuracy of this drawing, it should also be noted that the observational period in 1749-50 is situated in the solar maximum as shown in Figures 3 and 4, and thus it is expected to see larger spots around this time since the great sunspots are known to peak around and slightly after the solar maxima (Heath, 1994).

## 4. Conclusion

In this article, we examine sunspot drawings during 1749-50 in manuscripts of SKJZ. We found 15 sunspot drawings in SKJZ that include information not only of their number and location but also their area and shape. These manuscripts tell us that these observations were carried out at the Shogunate Observatory in Edo (current Tokyo) with telescopes covered by *zongrasu*s. These observations are located near the solar maximum and hence show relatively numerous sunspots and sometimes those with large area. We examined their group numbers, areas, and locations, while we need to admit the difference of variant manuscripts with one another. These sunspot drawings mostly fall in the dates without known observations registered in the latest database for the raw sunspot observations by Vaquero *et al.* (2016). Therefore, these sunspot drawings can contribute to reconstruct detailed solar activity in mid-18$^{th}$ century where we are in short of active observations of sunspots. Reconstructing the longer solar cycle in further detail will contribute not only to further understanding of the long-term solar activity (e.g. Clette *et al.*, 2014; Vaquero *et al.*, 2016; Svalgaard & Schatten, 2016; Usoskin, 2017), but also to further discussions on solar dynamo theory (e.g. Hotta *et al.*, 2016) or prediction of future solar activity (e.g. Svalgaard *et al.*, 2005; Iijima et al., 2017) in comparison with theoretical studies.

**Acknowledgement**

We acknowledge the support of Kyoto University's Supporting Program for the Interaction-based

Initiative Team Studies "Integrated study on human in space" (PI: H. Isobe), the Interdisciplinary Research Idea contest 2014 by the Center for the Promotion of Interdisciplinary Education and Research, the "UCHUGAKU" project of the Unit of Synergetic Studies for Space, the Exploratory and Mission Research Projects of the Research Institute for Sustainable Humanosphere (PI: H. Isobe), and SPIRITS 2017 (PI: Y. Kano) of Kyoto University. This work was also encouraged by a Grant-in-Aid from the Ministry of Education, Culture, Sports, Science and Technology of Japan, Grant Number JP15H05816 (PI: S. Yoden), JP16H03955 (PI: K. Shibata), JP16K17671 (PI: S. Toriumi), JP16K17655 (PI: H. Hotta), JP16H01169 (PI: H. Hotta), and JP15H05814 (PI: K. Ichimoto), and Grant-in-Aid for JSPS Research Fellow JP17J06954 (PI: H. Hayakawa). We especially thank the Tohoku University Library for their permission of researches and reproductions of the original manuscripts. We also thank SIDC for providing data on sunspot number and group number of sunspots, Dr. C. J. Schrijver for helpful advice on "zonglas" in modern Dutch, Dr. M. C. Werner for his helpful advice on the spelling variation of "Staudach" and "Staudacher", and Dr. Y. Hanaoka for his help in aligning the solar coordinates on the sunspot drawings.

**Appendix 1: References of Source Documents**

*Tokugawa Jikki*: Narushima Motonao et al., *Tokugawa Jikki* (K. Kuroita, ed.), v.9 (Tokyo: Yoshikawa Kobunkan), 1966

*Rekisan Zensho*: MS ni05-01614 in the Waseda University Library.

MS/K: MS 8-21318-1 in Kano Library of Tohoku University Library

MS/O: MS 911-17799 in Okamoto Library of Tohoku University Library

**Appendix 2: Estimating Areas and Locations of Sunspots**

In order to estimate areas and locations of sunspots, we use the following procedures.

**1. Circle Fitting:**

First, we cut out each sunspot image (S2 ~ S15) from the two original manuscripts and applied elliptical fitting to each spot image after the background trend was subtracted by the method in Section 2.2 of Toriumi *et al*. (2014). Then, we reformed the spot image so that the ellipse becomes a circle, and determined the center position and radius of the solar disc.

**2. Extraction of sunspots:**

In order to extract sunspots, we made a binary map of each image, leaving only sunspots, and labeled them with numbers. The intensity threshold that we used for each image is

$$Threshold = median - 3\sigma$$

### 3. Calculation of Sunspot Coordinates:

The north-south direction of the Sun observed from the earth does not match the vertical orientation of the image because of the inclination of the solar rotation axis with respect to the plane of revolution of the Earth (P angle). Therefore, we calculated the P angle at that time from the observational date in Table 1, and rotated the image such that the solar north-south comes to the image's vertical. Then, we calculated the coordinates on the solar surface considering the inclination of the solar equator to the ecliptic (B0 angle). In order to calculate these angles, we used "get_sun.pro" procedure of SSWIDL, which is based on the method in Meeus (1988).

### 4. Calculation of Projected Area of Sunspots

We calculated the projected are of each sunspot with the formula.

$$projected\ area = \frac{(\text{number of pixels composed sunspot})}{\pi R^2}$$

Here, $R$ means the radius of the corrected circle defined by the circle fitting. The unit of radius is the number of pixel in the reproduced figures.

### 5. Calculation of Corrected Area of Sunspots:

We projected each sunspot patch on a planar projection system in which we look down the spot vertically from above. We used "Sanson Projection" for the projection system. Here, the area per pixel is equal everywhere, which is about $1.474 \times 10^6 [km^2] \sim 0.484 [mhs]$. The area of the sunspot was derived from counting the number of the pixels defined as sunspot in the projected image and converting it into the unit of msh.

Table 1: Sunspot observations in SKJZ with their date and group number of sunspots in comparison with raw group counts (RGC) by Vaquero *et al.* (2016). We categorize sunspots within 15° within the same group. Note that S1 contains observations of four timings in the same drawing and we could not distinguish them from those in other timing, as further detail was not provided in this drawing in either of variants.

| ID | Year | Month | Date | MS/K | MS/O | RGC | Reference |
|---|---|---|---|---|---|---|---|
| S1 | 1749 | Aug. | 7 | | | N/A | f.9a |
| S1 | 1749 | Aug. | 8 | | | N/A | f.9a |
| S2 | 1749 | Oct. | 12 | 5 | 5 | 5 | f.9b |
| S3 | 1749 | Oct. | 15 | 6 | 6 | N/A | f.9b |
| S4 | 1749 | Oct. | 17 | 2 | 2 | N/A | f.9b |
| S5 | 1749 | Oct. | 19 | 2 | 2 | N/A | f.9b |
| S6 | 1749 | Oct. | 21 | 1 | 1 | N/A | f.9b |
| S7 | 1749 | Oct. | 23 | 1 | 1 | N/A | f.9b |
| S8 | 1749 | Nov. | 2 | 4 | 4 | 6 | f.9b |
| S9 | 1749 | Nov. | 3 | 2 | 2 | N/A | f.9b |
| S10 | 1749 | Nov. | 8 | 4 | 4 | N/A | f.9b |
| S11 | 1749 | Nov. | 14 | 1 | 1 | N/A | f.10a |
| S12 | 1749 | Nov. | 30 | 2 | 2 | N/A | f.10a |
| S13 | 1749 | Dec. | 2 | 2 | 4 | N/A | f.10a |
| S14 | 1749 | Dec. | 4 | 1 | 1 | N/A | f.10a |
| S15 | 1750 | Jan. | 11 | 2 | 2 | 5 | f.10a |

Table 2a: Area and location of sunspot groups in MS/K.

| MS/K | Projected Area [$msh$] | Corrected Area [$msh$] | Longitude | Latitude |
| --- | --- | --- | --- | --- |
| S02A_E01 | 426 | 260 | 9.5 | -25.8 |
| S02A_E02 | 403 | 220 | -3.9 | -15.9 |
| S02A_E03 | 892 | 460 | -10.9 | 13.7 |
| S02A_E04 | 747 | 430 | -28.6 | 16.9 |
| S02A_E05 | 683 | 400 | 12.7 | 34.1 |
| S03A_E01 | 434 | 290 | 24.2 | -29.6 |
| S03A_E02 | 366 | 210 | 6.9 | -22.3 |
| S03A_E03 | 318 | 190 | 28.8 | -8.5 |
| S03A_E04 | 695 | 470 | 43.1 | 4.1 |
| S03A_E05 | 917 | 500 | 23.5 | 9.6 |
| S03A_E06 | 637 | 540 | -54.1 | 7 |
| S03A_E07 | 1632 | 1140 | -43.7 | 12.4 |
| S04A_E01 | 2852 | 2010 | -45.4 | -3 |
| S04A_E02 | 1403 | 1150 | 52.7 | 25.4 |
| S05A_E01 | 1696 | 860 | 11.8 | -3.9 |
| S05A_E02 | 990 | 1480 | 70.7 | 0.1 |
| S06A_E01 | 3478 | 1990 | 29.9 | 3.2 |
| S07A_E01 | 3705 | 4670 | 65.5 | 2.4 |
| S08A_E01 | 1878 | 1060 | 25.4 | -7.8 |
| S08A_E02 | 705 | 370 | -16.4 | -7.3 |
| S08A_E03 | 634 | 320 | -12.7 | -5.4 |
| S08A_E04 | 978 | 540 | -19.7 | 19.6 |
| S08A_E05 | 1387 | 740 | -15.6 | 21.4 |
| S08A_E06 | 1327 | 930 | 18.1 | 45.8 |
| S09A_E01 | 1383 | 790 | 24.7 | -13.9 |
| S09A_E02 | 1725 | 1020 | 25 | 27.1 |
| S10A_E01 | 1328 | 710 | 12.3 | -12.5 |
| S10A_E02 | 1193 | 640 | 16.1 | -9.1 |
| S10A_E03 | 711 | 510 | 45.1 | -4.3 |
| S10A_E04 | 646 | 390 | -32 | 10.4 |
| S10A_E05 | 735 | 610 | 51.5 | 18.8 |
| S10A_E06 | 693 | 470 | 40 | 19.8 |
| S10A_E07 | 826 | 680 | 49.3 | 27 |
| S11A_E01 | 1142 | 700 | -33.3 | -10.6 |
| S12A_E01 | 2165 | 1240 | 28.9 | -1.6 |
| S12A_E02 | 1518 | 960 | 34 | 18.2 |
| S13A_E01 | 1629 | 1360 | 50.4 | -19.3 |
| S13A_E02 | 820 | 670 | 47.9 | 25.8 |
| S14A_E01 | 876 | 890 | 60.3 | 16.9 |
| S15A_E01 | 647 | 400 | -36.5 | -2.7 |
| S15A_E02 | 1765 | 1020 | -27 | 11.5 |

Table 2b: Area and location of sunspot groups in MS/O.

| MS/O | Projected Area [msh] | Corrected Area [msh] | Longitude | Latitude |
| --- | --- | --- | --- | --- |
| S02B_E01 | 462 | 280 | 9.7 | -25.6 |
| S02B_E02 | 341 | 190 | -4.2 | -16.7 |
| S02B_E03 | 645 | 330 | -11.3 | 13.2 |
| S02B_E04 | 478 | 280 | -29.4 | 16.9 |
| S02B_E05 | 532 | 320 | 12 | 34.4 |
| S03B_E01 | 555 | 380 | 26 | -29.7 |
| S03B_E02 | 355 | 210 | 8.7 | -23.5 |
| S03B_E03 | 258 | 160 | 30.7 | -9.2 |
| S03B_E04 | 453 | 330 | 46.4 | 2.8 |
| S03B_E05 | 559 | 310 | 25.4 | 10.3 |
| S03B_E06 | 486 | 440 | -55.8 | 7.8 |
| S03B_E07 | 1475 | 1020 | -42.2 | 12.7 |
| S04B_E01 | 2153 | 1590 | -46.9 | -1.2 |
| S04B_E02 | 1194 | 1100 | 56.6 | 24.9 |
| S05B_E01 | 1441 | 750 | 12.5 | -2.9 |
| S05B_E02 | 939 | 1240 | 67.7 | 2 |
| S06B_E01 | 3548 | 2070 | 30.4 | 3.8 |
| S07B_E01 | 2462 | 3450 | 68.2 | 1.7 |
| S08B_E01 | 890 | 510 | 28 | -9 |
| S08B_E02 | 605 | 310 | -14.1 | -8.2 |
| S08B_E03 | 878 | 450 | -10.9 | -5.8 |
| S08B_E04 | 736 | 390 | -16 | 18.2 |
| S08B_E05 | 981 | 520 | -11.4 | 20.4 |
| S08B_E06 | 1029 | 690 | 23.5 | 40.4 |
| S09B_E01 | 1096 | 650 | 25.4 | -15.5 |
| S09B_E02 | 1623 | 960 | 26.4 | 24.6 |
| S10B_E01 | 911 | 470 | 11.7 | -11.5 |
| S10B_E02 | 1406 | 730 | 16.2 | -8.3 |
| S10B_E03 | 784 | 570 | 47.5 | -4 |
| S10B_E04 | 952 | 520 | -30.4 | 13.3 |
| S10B_E05 | 708 | 650 | 57.2 | 18.9 |
| S10B_E06 | 483 | 340 | 43.8 | 21.1 |
| S10B_E07 | 1013 | 920 | 54.2 | 28.7 |
| S11B_E01 | 693 | 430 | -35 | -9.6 |
| S12B_E01 | 1310 | 750 | 30.1 | 0.6 |
| S12B_E02 | 1168 | 760 | 36.6 | 19.4 |
| S13B_E01 | 1140 | 1090 | 55.5 | -19.3 |
| S13B_E02 | 865 | 890 | 57.2 | 28.4 |
| S13B_E03 | 604 | 300 | 0.1 | 3.8 |
| S13B_E04 | 1755 | 2040 | 51.4 | 45.8 |
| S14B_E01 | 532 | 570 | 61.7 | 17.9 |
| S15B_E01 | 928 | 580 | -34.8 | -0.9 |
| S15B_E02 | 1446 | 870 | -26.3 | 14.5 |

Figure 1a: f. 9a of MS/K

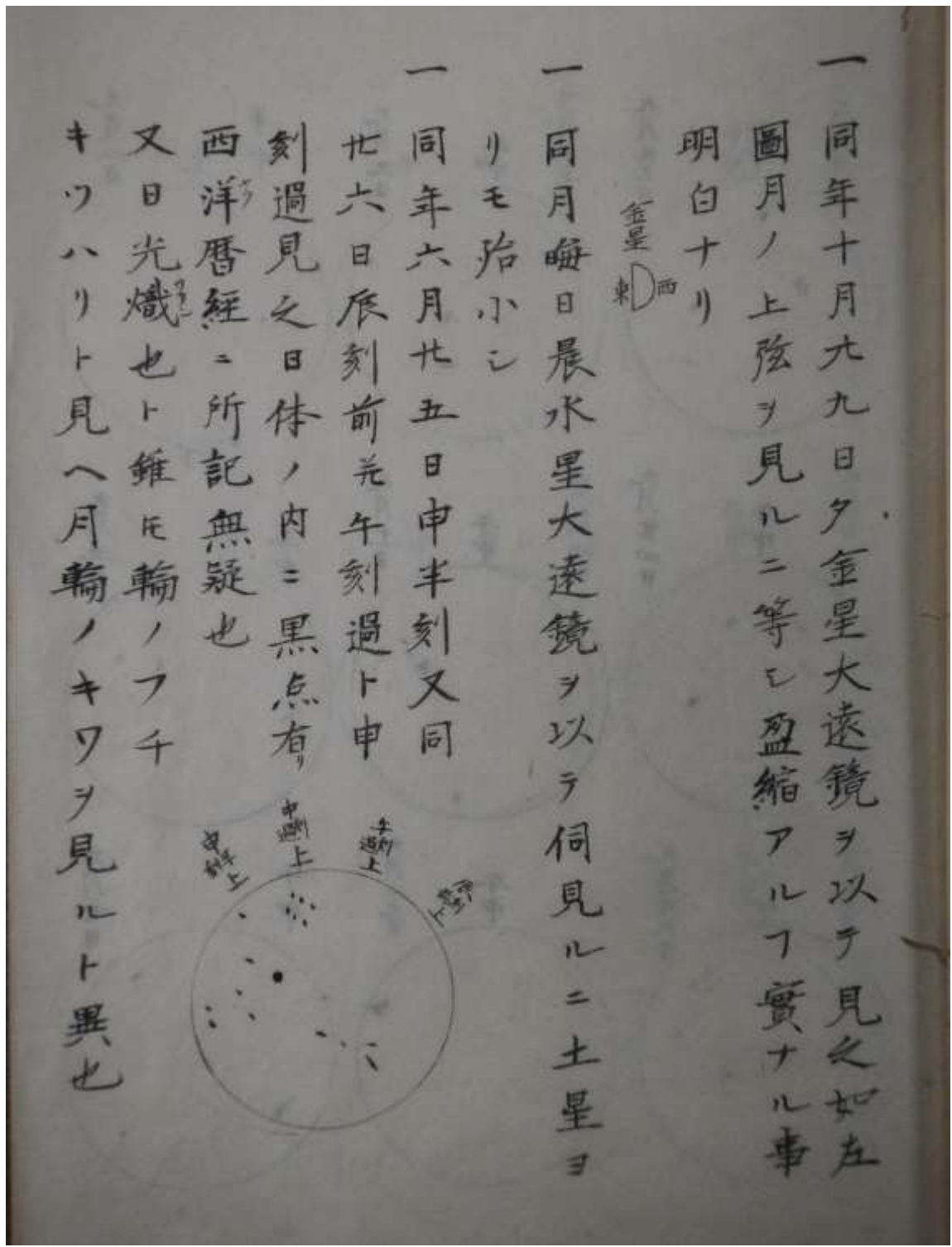

一 同年十月九日夕金星大遠鏡ヲ以テ見ル之如左
圖月ノ上弦ヲ見ルニ等シ盈縮アルフ實ナル事
明白ナリ

一 同月晦日晨水星大遠鏡ヲ以テ伺見ルニ土星ヨ
リモ始小シ

一 同年六月廿五日申半刻又同
廿六日辰刻前先午刻過ト申
刻過見之日体ノ内ニ黒点有
西洋暦経ニ所記無疑也
又日光熾也ト雖モ輪ノフチ
キワハリト見ヘ月輪ノキワヲ見ルト異也

Figure 1b: f. 9b of MS/K

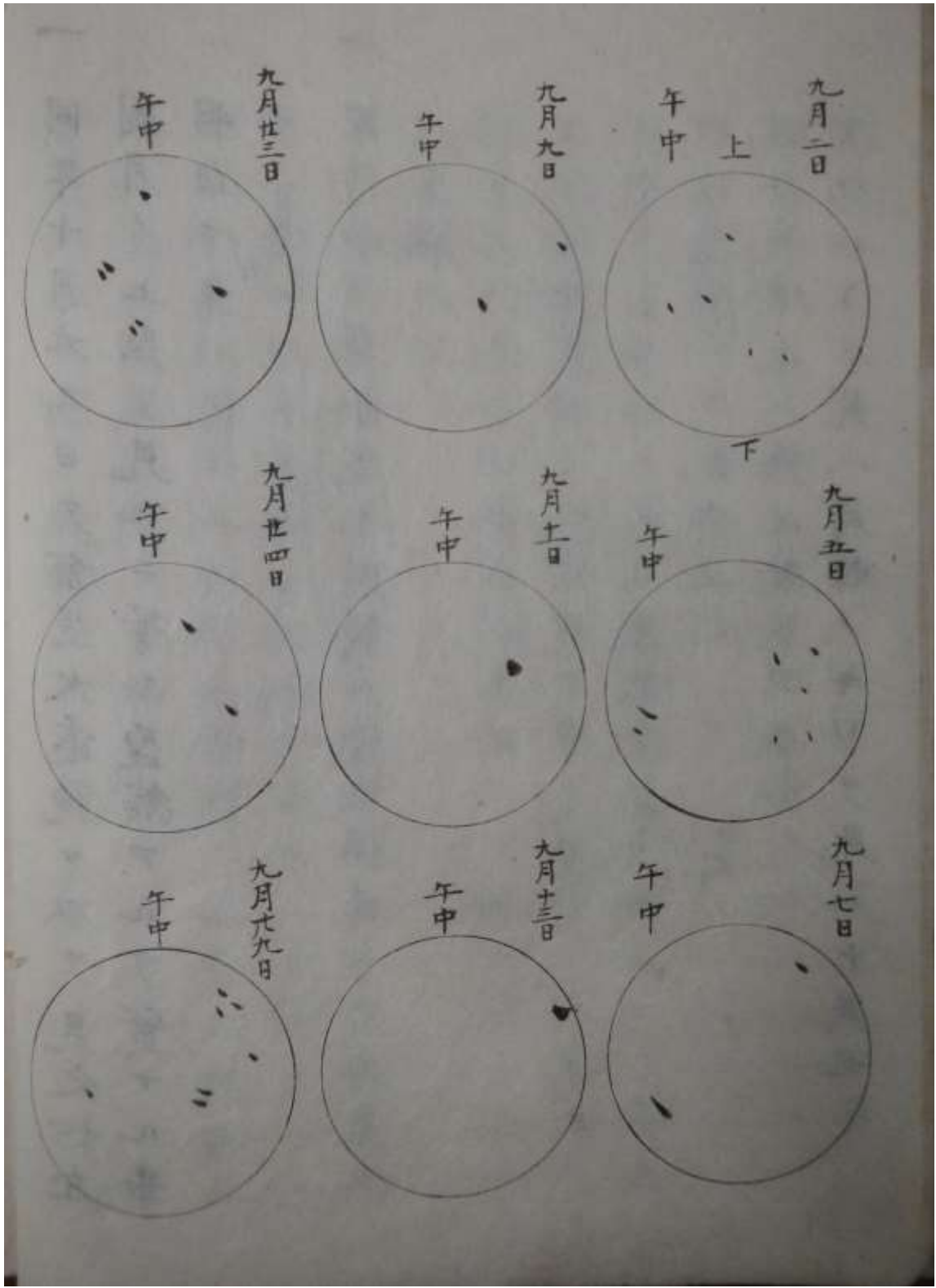

Figure 1c: f. 10a of MS/K

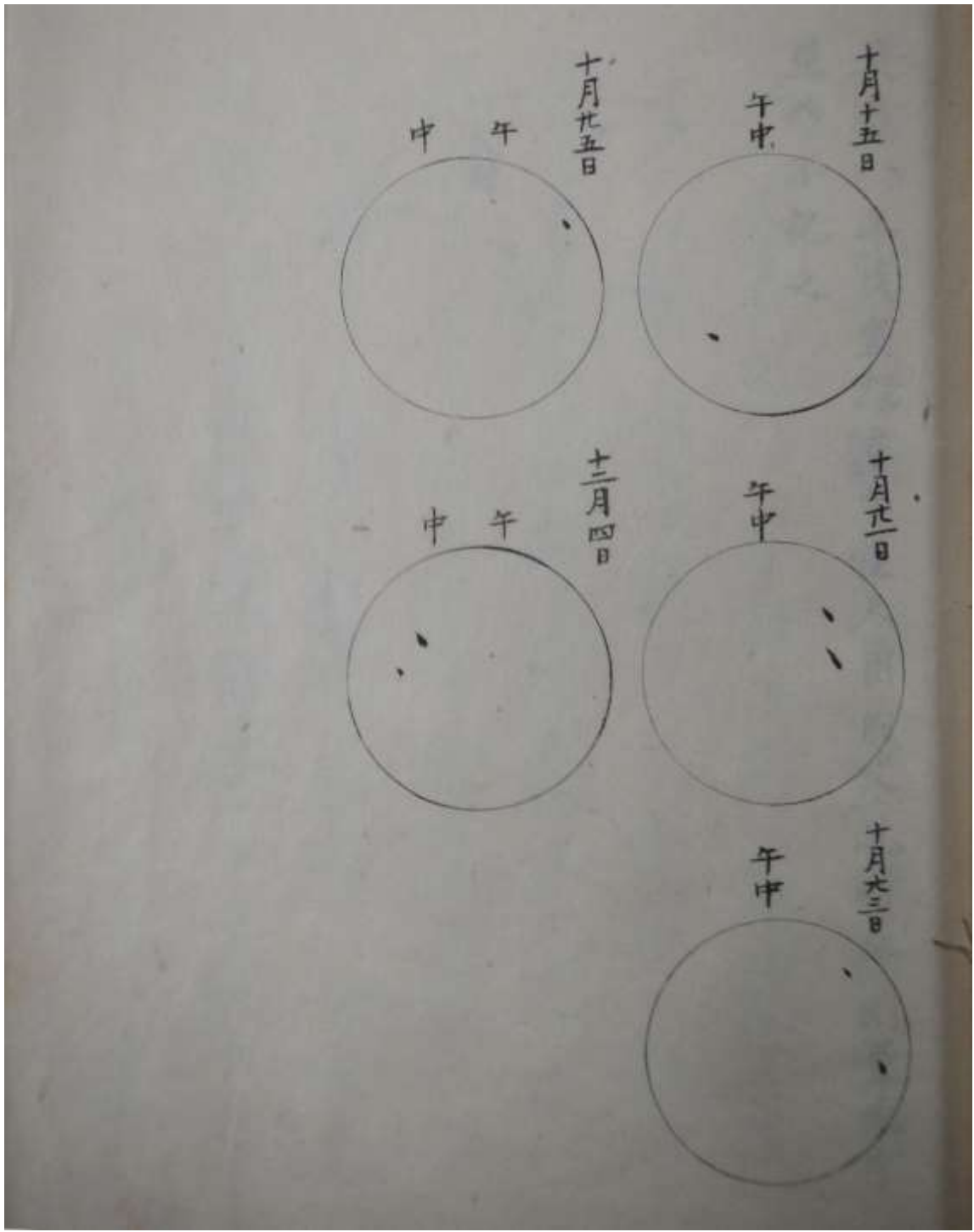

Figure 2a: f. 9a of MS/O

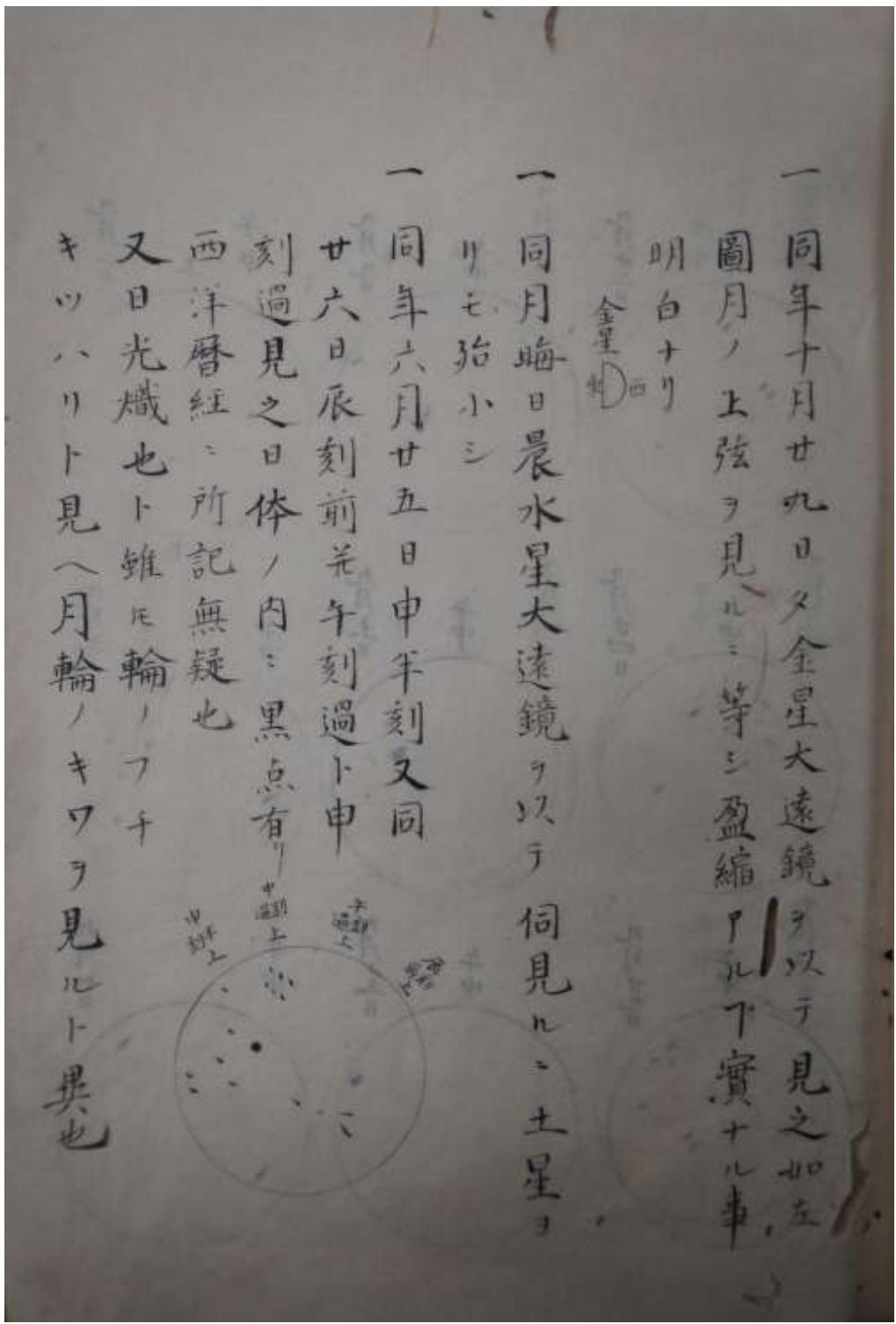

一同年十月廿九日夕金星大遠鏡ヲ以テ見之如左
圖月ノ上弦ヲ見ルニ等シ盈縮アルト實ナル事
明白ナリ
金星

一同月晦日晨水星大遠鏡ヲ以テ伺見ルニ土星
ヲ始小シ

一同年六月廿五日申午刻又同
廿六日辰刻前羗午刻過ト申
刻過見之日体ノ内ニ黒点有リ
西洋暦経ニ所記無疑也
又日光熾也ト雖ニ輪ノフチ
キツハリト見ヘ月輪ノキワヲ見ルト異也

Figure 2b: f. 9b of MS/O

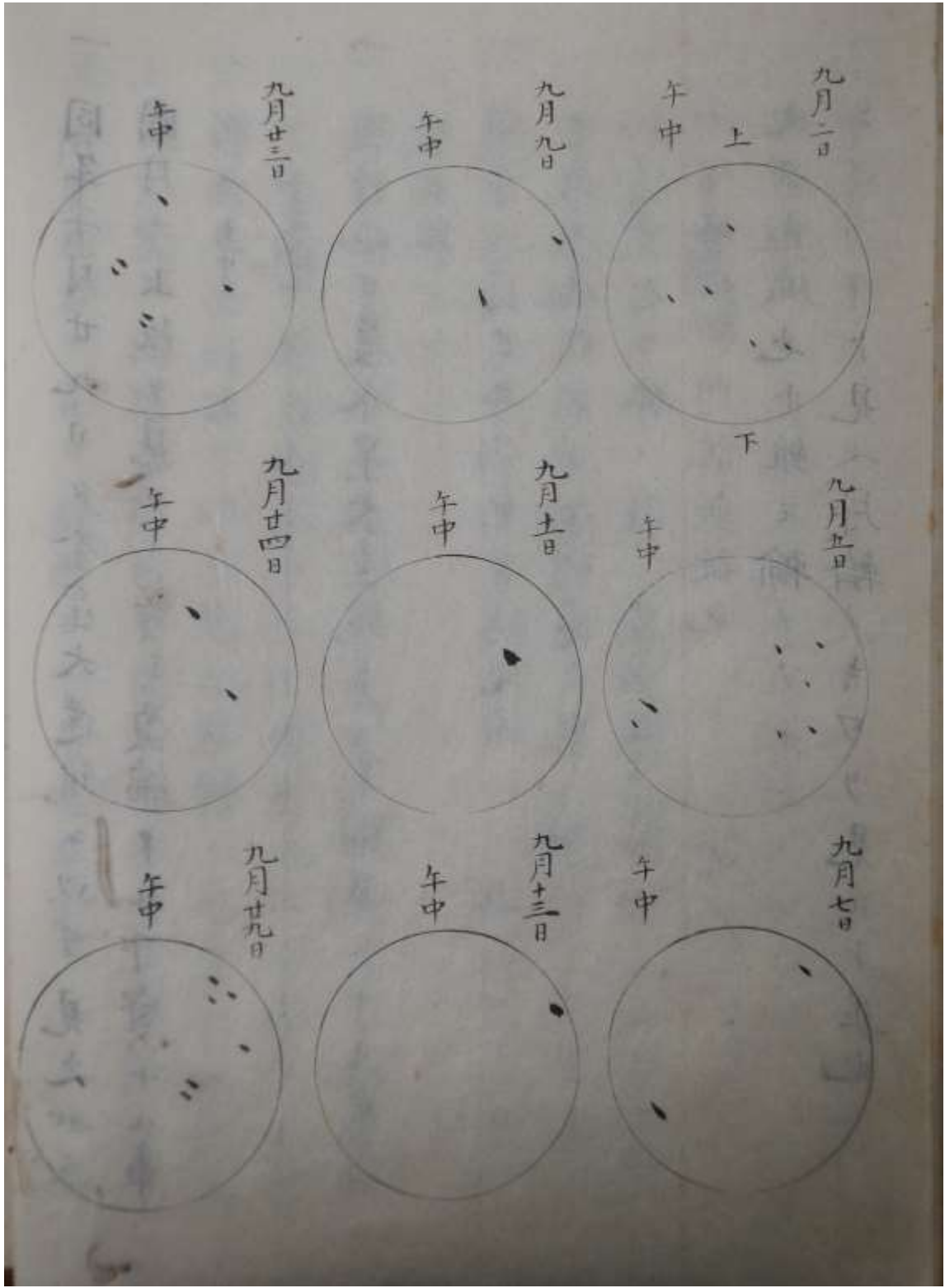

Figure 2c: f. 10a of MS/O

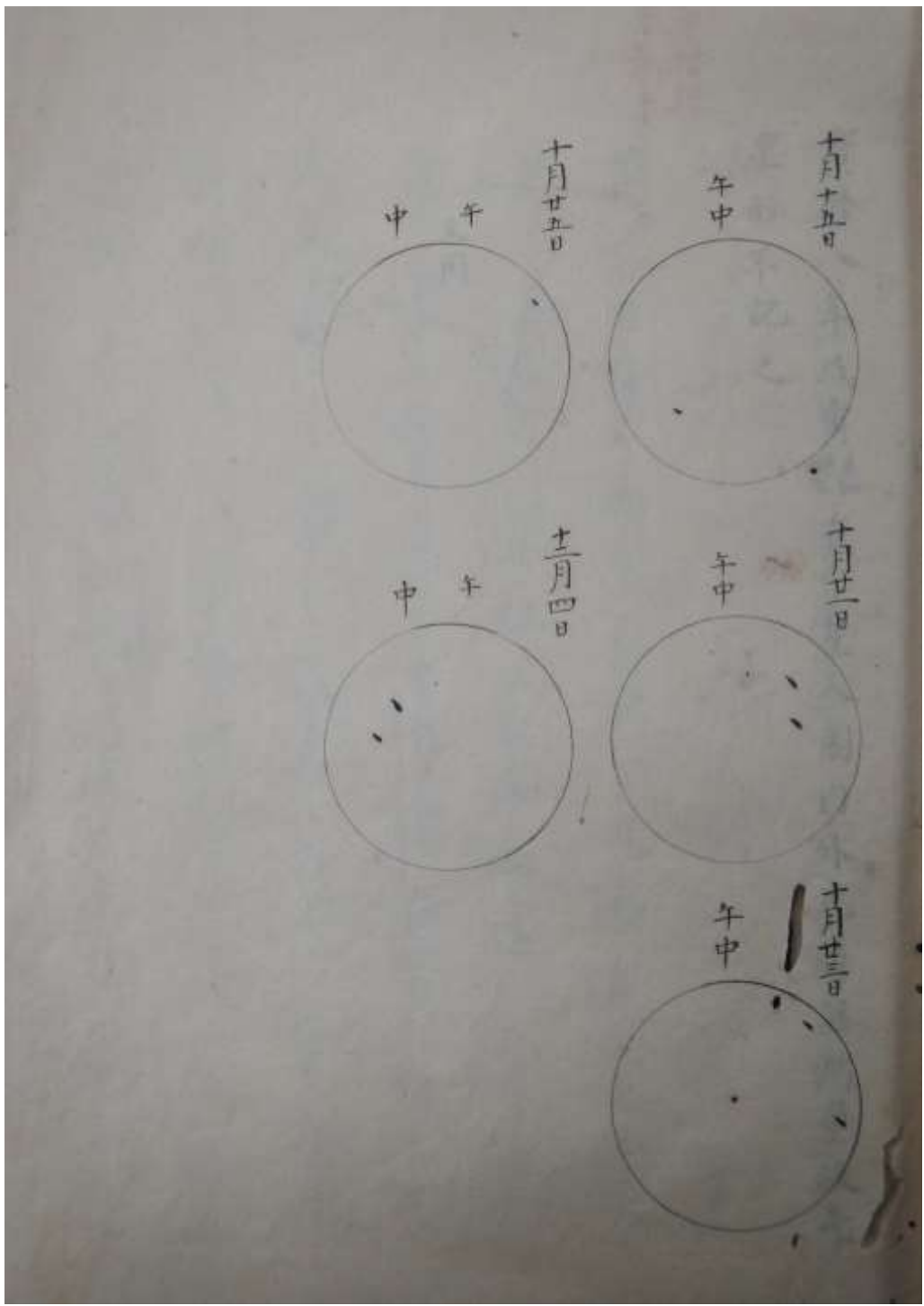

Figure 3: Identification of sunspot groups:

(a) Sunspot groups in MS/K

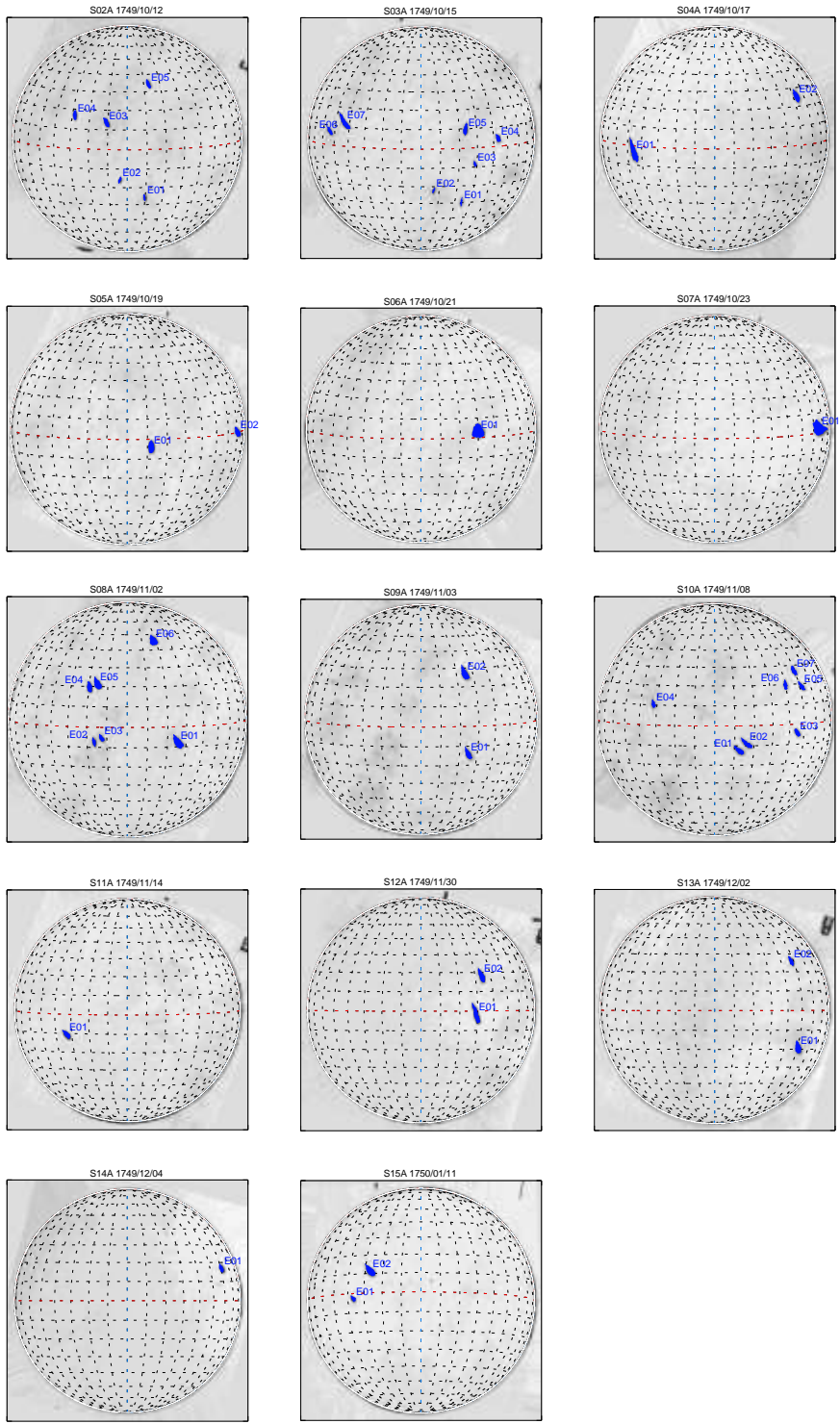

(b) Sunspot groups in MS/O

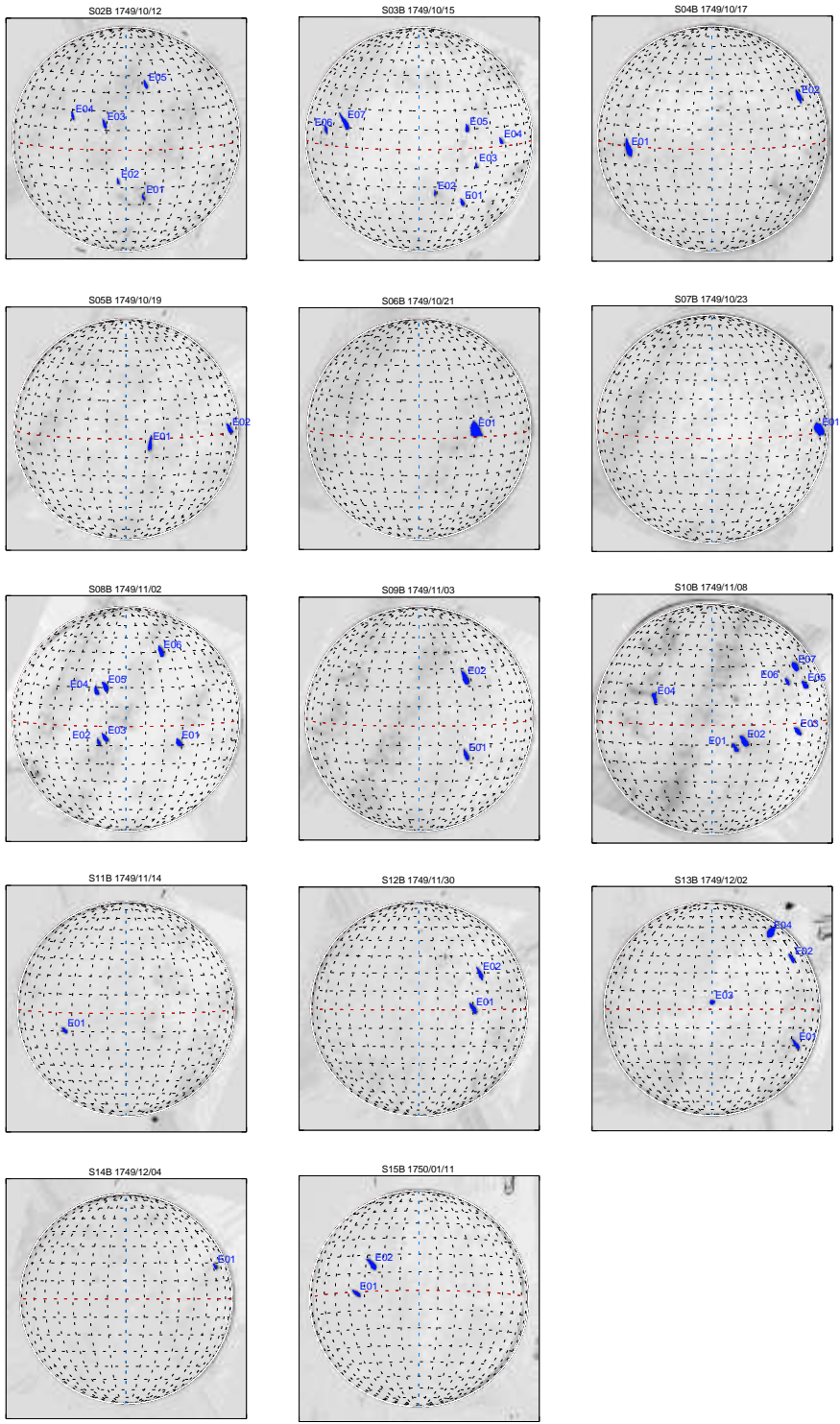

Figure 4: Sunspot drawings in SKJZ within the index of yearly sunspot number by Clette *et al.* (2014)

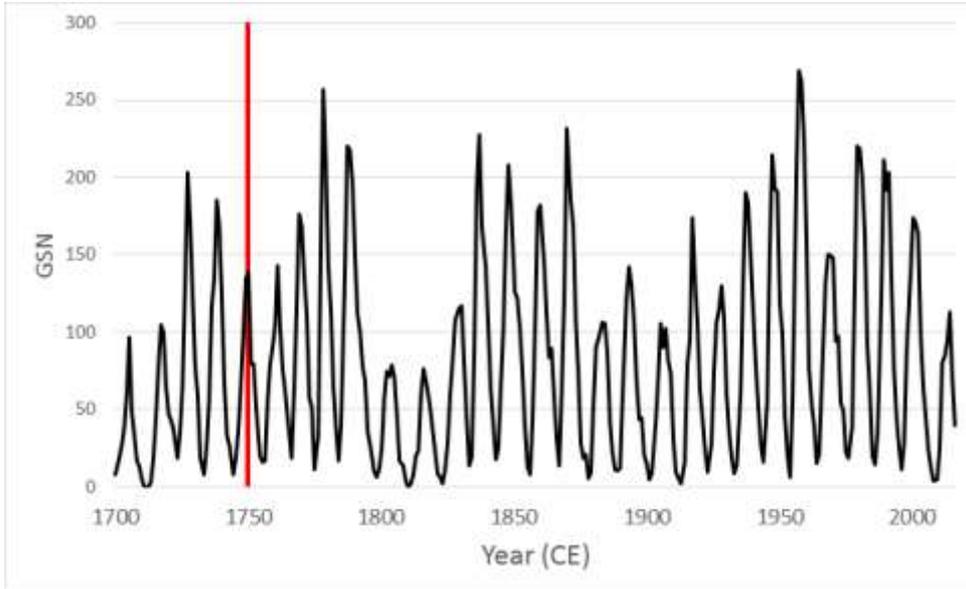

Figure 5: Sunspot drawings in SKJZ within the index of monthly sunspot number by Clette *et al.* (2014). The coverage of blue color shows where the sunspot drawings in SKJZ are situated.

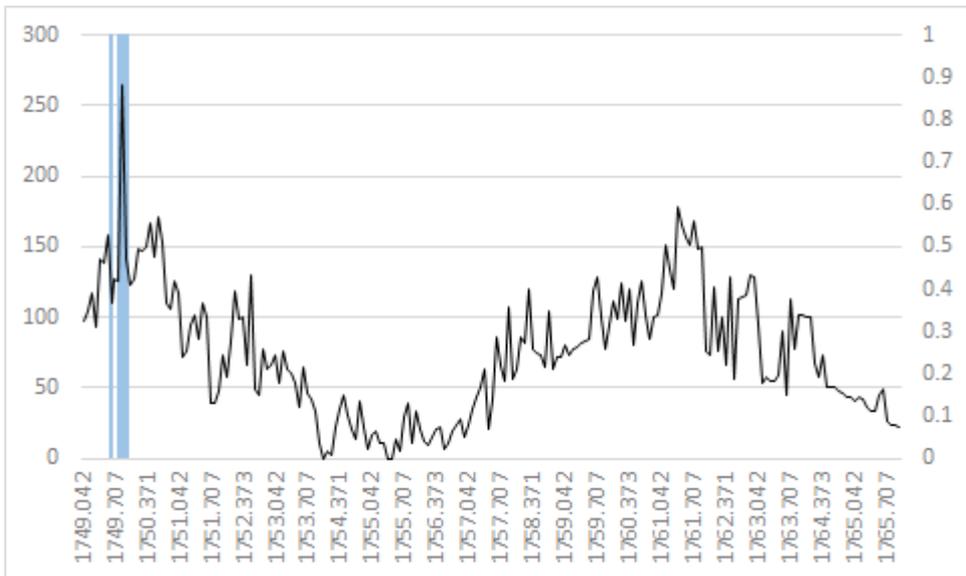

Figure 6: The raw group count of sunspots in the SKJZ in comparison with those by Vaquero *et al.* (2016).

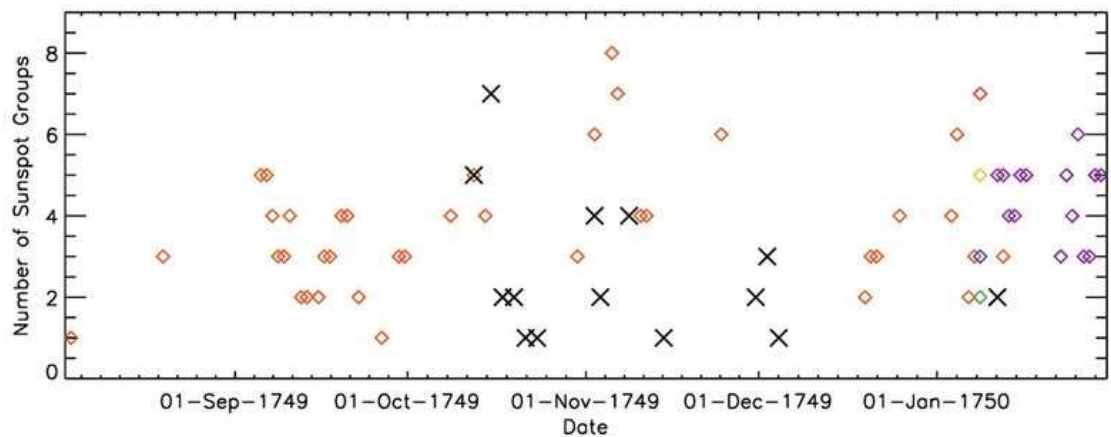

Crosses: raw group count of sunspots in SKJZ

Blue squares: Cassini in Paris

Orange squares: Staudach[5] in Nuremberg

Red squares: Zanotti in Bononia

Purple squares: Hagen in Berlin

Yellow-green square: Danquier in Paris

Green squares: squares: Messier in Paris

---

[5] His name is sometimes spelled "Staudach" (e.g. Clette *et al.*, 2014; Svalgaard and Schatten, 2016; Vaquero *et al.*, 2016; Svalgaard, 2017) and sometimes "Staudacher" (e.g. Arlt, 2008; Arlt, 2009; Arlt and Fröhlich, 2012; Pavai *et al.*, 2016). It is known that Staudach himself signed his own name as "Staudach" (e.g., Figure 1 of Arlt 2008) but was also known as "Staudacher", as explained in the correspondence by Gustav Spörer cited in Wolf (1887, pp.357-358). Note that both Spörer and Wolf were professional astronomers publishing in German in the nineteenth century. Therefore, it should be noted that either spelling is possible. We adopt "Staudach" here to keep our article consistent with Vaquero *et al.* (2016).

Figure 7: Distribution of projected area (above) and corrected area (below) of sunspots with manuscript variation. Basically, upper end of variation indicates the area of MS/K and lower end is MS/O. Their average value is plotted using square markers. On the other hand, triangle markers are used when MS/K is smaller than MS/O. Each color corresponds with the same sunspot in different dates. Element number (e.g. E01, E02, etc.) is increased from left side. S13_E03 and S13_04 don't have variation because these spots are not found in MS/K.

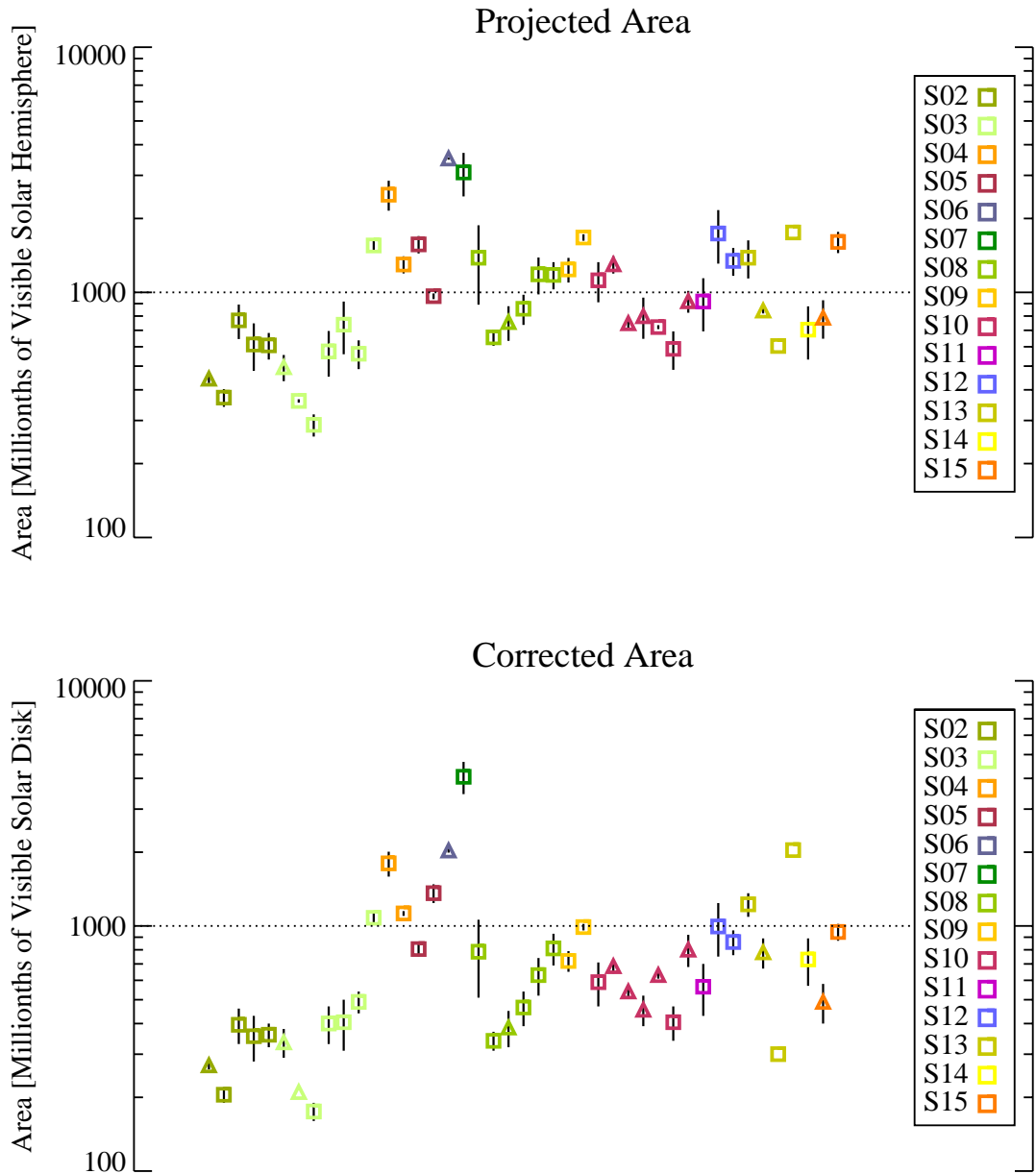

Figure 8: Comparison of histograms of distribution of projected area and corrected area of sunspots in MS/K and MS/O.

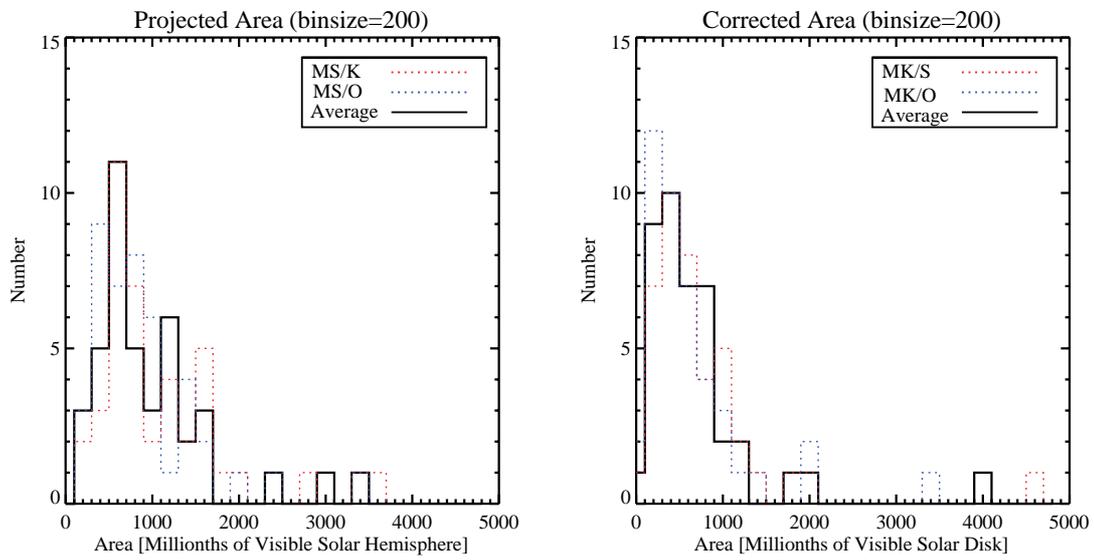